\documentclass[aps,pre,showpacs,superscriptaddress]{revtex4}

\usepackage[dvips]{graphicx}
\usepackage{mathrsfs}
\usepackage{amsmath,amssymb}
\usepackage{hhline}
\usepackage{dcolumn} % Align table columns on decimal point
\usepackage{bm} % bold math
\usepackage[dvips]{color}

\newcommand{\e}{\mbox{\rm e}}
\renewcommand{\d}{\mbox{\rm d}}
\newcommand{\Li}{\mbox{\rm Li}}

\begin{document}

\title{Efficiency of prompt quarantine measures on a susceptible-infected-removed model in networks}

\author{Takehisa Hasegawa}
\email{takehisa.hasegawa.sci@vc.ibaraki.ac.jp}
\affiliation{%
Department of Mathematics and Informatics, 
Ibaraki University, 
2-1-1, Bunkyo, Mito, 310-8512, Japan
}%
\author{Koji Nemoto}
\email{nemoto@statphys.sci.hokudai.ac.jp}
\affiliation{%
Department of Physics, Hokkaido University,
Kita 10 Nishi 8, Kita-ku, Sapporo, Hokkaido, 060-0810, Japan
}%

\begin{abstract}
This study focuses on investigating the manner in which a prompt quarantine measure suppresses epidemics in networks. A simple and ideal quarantine measure is considered in which an individual is detected with a probability immediately after it becomes infected and the detected one and its neighbors are promptly isolated. The efficiency of this quarantine in suppressing a susceptible-infected-removed (SIR) model is tested in random graphs and uncorrelated scale-free networks. Monte Carlo simulations are used to show that the prompt quarantine measure outperforms random and acquaintance preventive vaccination schemes in terms of reducing the number of infected individuals. The epidemic threshold for the SIR model is analytically derived under the quarantine measure, and the theoretical findings indicate that prompt executions of quarantines are highly effective in containing epidemics. Even if infected individuals are detected with a very low probability, the SIR model under a prompt quarantine measure has finite epidemic thresholds in fat-tailed scale-free networks in which an infected individual can always cause an outbreak of a finite relative size without any measure. The numerical simulations also demonstrate that the present quarantine measure is effective in suppressing epidemics in real networks.
\end{abstract}

%\keywords{}

%\maketitle must follow title, authors, abstract, \pacs, and \words
\maketitle

%%%%%%%%%%%%%%%%%%%%%%%%%%%%%%%%%%%

\section{Introduction}

Recently, several studies were devoted to examining the spread of epidemics on networks in which nodes represent individuals and edges represent their social or sexual relationships through which an infectious disease spreads (as shown in review \cite{pastor2014epidemic} and references therein). Theoretical studies for epidemiological models demonstrated that infectious diseases could spread very easily in highly heterogeneous networks \cite{newman2003structure,barrat2008dynamical}. Specifically, two fundamental epidemic models, namely the susceptible-infected-removed (SIR) model \cite{kermack1927contribution} and the susceptible-infected-susceptible model \cite{anderson1992infectious}, exhibit outbreaks of finite relative sizes with an infinitesimal infection rate if the underlying network is fat-tailed scale-free such that the degree distribution obeys $p_k \propto k^{-\gamma}$, with $\gamma \le 3$ \cite{pastor2001epidemic,moreno2002epidemic}.

In order to contain epidemics, several control measures were proposed that utilize network information. Epidemics can be suppressed by effective vaccination schemes such as the target vaccination \cite{holme2004efficient}, the acquaintance vaccination \cite{cohen2003efficient, gallos2007improving}, the PageRank-based vaccination \cite{miller2007effective}, and the graph partitioning vaccination \cite{chen2008finding}. Theoretically, the above vaccination schemes succeed in containing epidemics in which a network is highly heterogeneous although these vaccination schemes are considered as a preventive measure wherein it is necessary to complete vaccinations prior to the appearance of an infectious disease in a network.

With respect to postoutbreak strategies, previous studies examined local control measures in which susceptible individuals who were in contact with an infected individual are vaccinated or isolated \cite{dybiec2004controlling,dybiec2005optimising,takeuchi2006effectiveness,shaban2008networks,oles2012understanding,karp2014improving,xu2014comparative}.
Dybiec et al. \cite{dybiec2004controlling,dybiec2005optimising} considered a spatial epidemic model in a situation in which individuals can be infectious prior to exhibiting symptoms (and therefore prior to detection), and a local control measure is probabilistically applied in a neighborhood centered around a detected infectious individual.
The results indicated the optimal radius necessary for the aforementioned type of a control neighborhood to contain epidemics in terms of economic costs associated with disease and treatment.
Takeuchi and Yamamoto \cite{takeuchi2006effectiveness} studied a ring vaccination in which susceptible individuals who came in contact with infected ones were probabilistically vaccinated. The findings revealed that the ring vaccination scheme reduced  the infection rate and the number of vaccinated nodes becomes considerably small when compared to those in the preventive strategies. However, the basic reproduction number (and thus the epidemic threshold) remained equal to those of random preventive vaccination, and this failed to contain epidemics in a highly heterogeneous network unless almost all individuals were vaccinated.
There are also studies investigating dynamic reactions of individuals to the spread of epidemics \cite{bagnoli2007risk,lagorio2011quarantine,sahneh2012existence,wu2012impact,ruan2012epidemic,zhang2014suppression}, such as behavioral responses of individuals by reducing their contact rates \cite{zhang2014suppression}, based on the number of infected neighbors or by rewiring connections (i.e., disconnecting their connections to infected neighbors and reconnecting others) \cite{lagorio2011quarantine}.

In order to clarify the extent to which an ideal quarantine measure suppresses epidemics, the present study considers a simple case in which an individual is detected with a probability immediately after it becomes infected and the detected one and its neighbors are promptly quarantined. The efficiency of the prompt quarantine measure is numerically and analytically investigated to suppress SIR epidemics in typical networks in terms of the mean outbreak size, the epidemic threshold, and the occurrence probability of global outbreaks. The prompt quarantine measure is highly effective in containing epidemics, and it can theoretically eradicate epidemics in highly heterogeneous networks even when infected individuals are detected with a very small probability. The numerical simulations also indicate that the quarantine measure is effective in real networks.

\section{Model}

%%%%%%%%%%%%%%%%%%%%%%%%%%%%%%%%%%%
\begin{figure}
%[!b]
\begin{center}
\includegraphics[width=80mm]{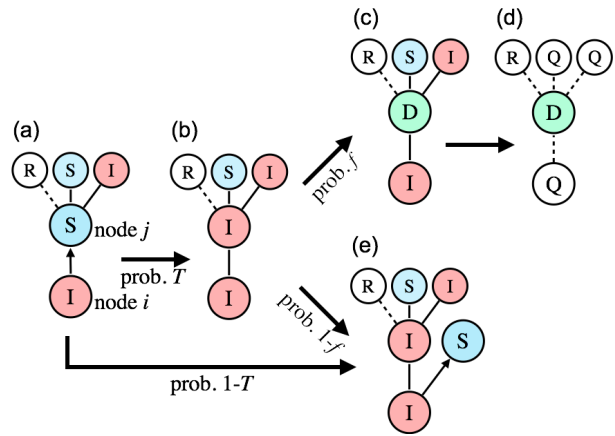}
\end{center}
\caption{
Transition rules of the SIR model with quarantine measures. (a) A node $j$ is randomly selected among the susceptible neighbors of  infected node $i$. (b) With probability $T$, disease is transmitted from $i$ to $j$. Immediately after (b), (c) this newly infected node $j$ is detected with probability $f$, and (d) node $j$ and its susceptible and infected neighbors (including node $i$) are promptly isolated. (e) If node $j$ is not detected, node $i$ tries to infect the next susceptible neighbor.
}
\label{fig:schematic}
\end{figure}
%%%%%%%%%%%%%%%%%%%%%%%%%%%%%%%%%%%

A discrete-time SIR model in a network is considered. For a given network with $N$ nodes, each node corresponds to one of the following three states: susceptible (S), infected (I), or removed (R). Any S node can be infected by contact with adjacent I nodes. An I node infects each of its S neighbors independently with probability $T$ and then spontaneously becomes R. A node that changes to the state R loses its capability to infect other nodes and does not change its state any further. The dynamics of the whole system is as follows: 
\begin{enumerate}
\item[(1)] 
Randomly select a node as a seed. As an initial configuration, all nodes except the seed are set to S, and the seed is set to I.
\item[(2)] 
Randomly select an I node $i$. Compile a new list of the S neighbors of node $i$. Randomly select a node from the list and change its state from S to I with probability $T$. Repeat this procedure until the list is empty, and then change the state of node $i$ to R. 
\item[(3)] Continue step (2) until I nodes cease to exist. That is, each node belongs to either S or R state in a final configuration.
\end{enumerate}

The SIR model placed on a network has an epidemic threshold $T_c$, in which an epidemic commencing from a seed terminates at an early stage for $T<T_c$, and a seed can cause a global outbreak (an outbreak of a finite relative size) for $T>T_c$. The order parameter $r$ defined by the mean fraction of the R nodes in final configurations is used to obtain $r=0$ for $T \le T_c$ and $r>0$ for $T > T_c$ in the limit $N \to \infty$. The epidemic threshold depends on the structure of the underlying network. With respect to uncorrelated networks with degree distribution $p_k$, the local tree approximation gives $T_c = \langle k \rangle/ \langle k(k-1) \rangle$ \cite{newman2002spread}, where $\langle \cdot \rangle$ represents the average of a quantity weighted by $p_k$. This indicates that it is considerably easier for a global outbreak to occur on heterogeneous networks when compared to homogeneous networks: $T_c=0$ for fat-tailed scale-free networks (SFNs) of $p_k \propto k^{-\gamma}$ with $2 < \gamma \le 3$, although $T_c = 1/\langle k \rangle > 0$ for random graphs (RGs) in which the degree distribution obeys $p_k \approx \langle k \rangle^k {\rm e}^{-\langle k \rangle}/k!$ with the same mean degree as that of the SFNs.

This is followed by introducing a prompt quarantine measure with respect to the SIR model. The proposed quarantine measure assumes that node $i$ can be detected (for example, by public health authorities) with a detection probability $f$ immediately after it becomes infected, and the detected node $i$ and its neighbors (except nodes already removed or quarantined) are promptly isolated. The detected and quarantined nodes lose the capability to infect others and to be further infected. It is also assumed that nodes already infected are cured by appropriate treatments when they are isolated. In order to incorporate this quarantine measure, an extended SIR model is considered by introducing the following additional states: detected (D) and quarantined (Q). 
The complete dynamics is modified as follows: 
\begin{enumerate}
\item[(2$^\prime$).] 
Randomly select an I node $i$. Compile a new list of the S neighbors of node $i$. Randomly select a node $j$ from the list (Fig.\ref{fig:schematic}(a)). With probability $T$, a disease is transmitted from node $i$ to node $j$; i.e., the state of $j$ is changed to I (Fig.\ref{fig:schematic}(b)). Immediately after that, change the state of $j$ to D with probability $f$ (Fig.\ref{fig:schematic}(c)). If node $j$ becomes D, then change the state of its S and I neighbors to Q (Fig.\ref{fig:schematic}(d)) and go to step (3). If node $j$ is not D, repeat the procedure until the list is empty (Fig.\ref{fig:schematic}(e)), and subsequently change the state of node $i$ to R.
\end{enumerate}
It should be noted that an I node attempts to infect each of its S neighbors, but this type of a process stops immediately when one of its neighbors becomes D.

%%%%%%%%%%%%%%%%%%%%%%%%%%%%%%%%%%%
\begin{figure}
%[!b]
\begin{center}
\includegraphics[width=80mm]{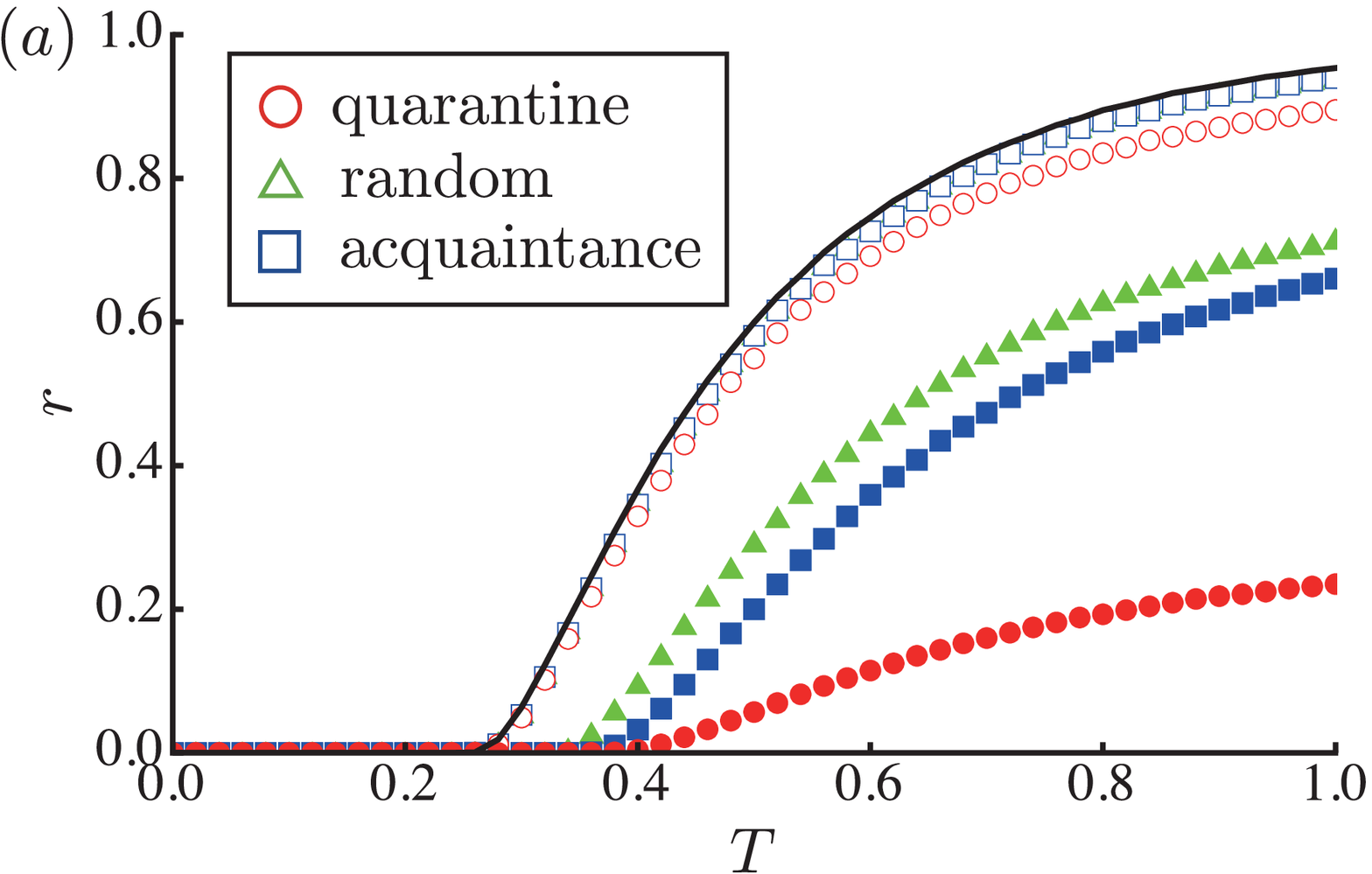}
\includegraphics[width=80mm]{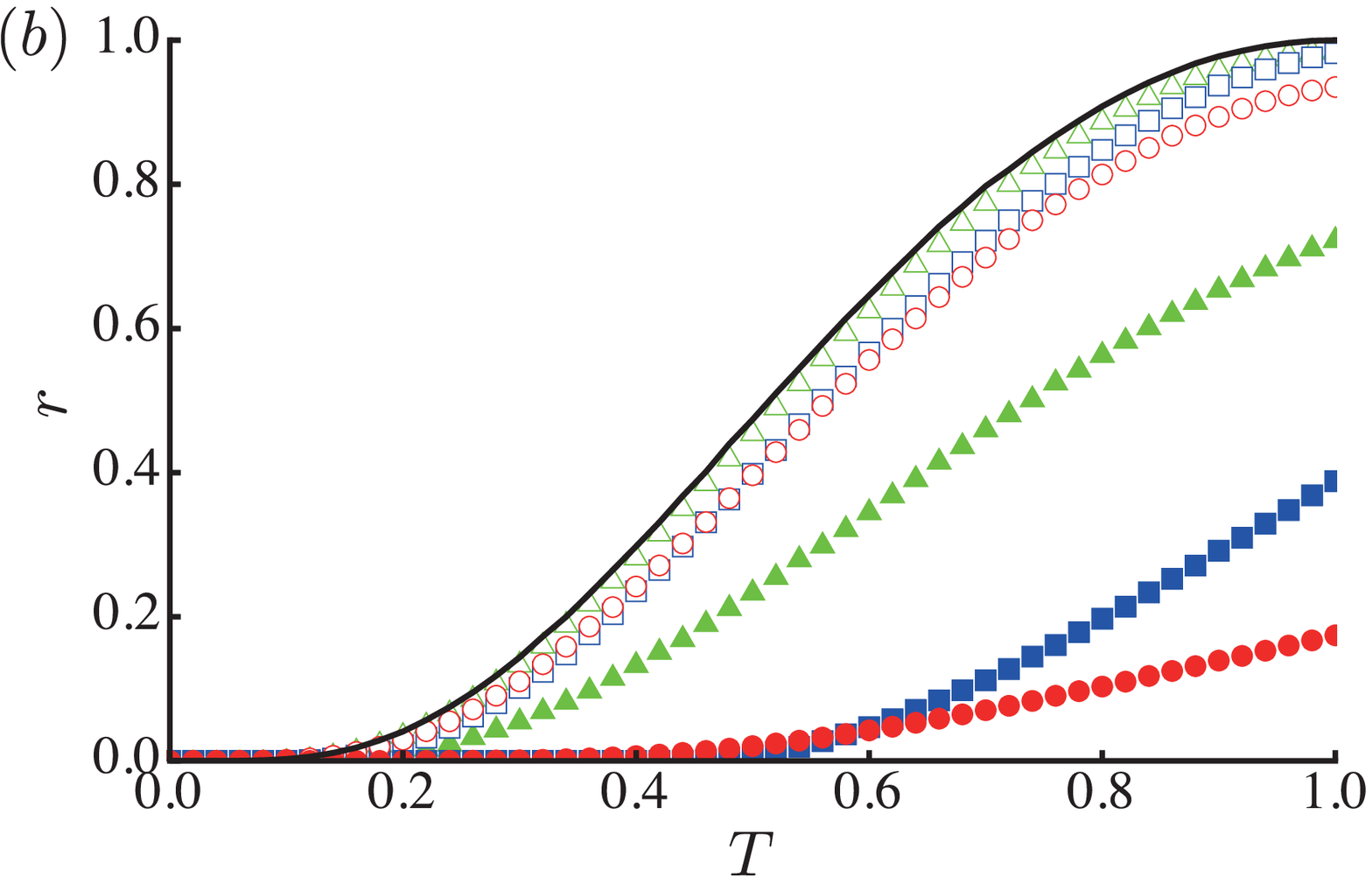}
\end{center}
\caption{
The order parameter $r$, as a function of $T$, for (a) the RGs and (b) the SFNs. Here $f$ is set to $f=0.01$ (open symbols) and 0.2 (full symbols). The red circles, green triangles, and blue squares represent the results for the quarantine measure, the random vaccination, and the acquaintance vaccination, respectively. The black line represents the result of the original SIR model. In (a), the open green triangles (random vaccination with $f=0.01$) fall behind the open blue squares (acquaintance vaccination with $f=0.01$).
}
\label{fig:orderparameter}
\end{figure}
%%%%%%%%%%%%%%%%%%%%%%%%%%%%%%%%%%%

%%%%%%%%%%%%%%%%%%%%%%%%%%%%%%%%%%%
\begin{figure}
%[!b]
\begin{center}
\includegraphics[width=80mm]{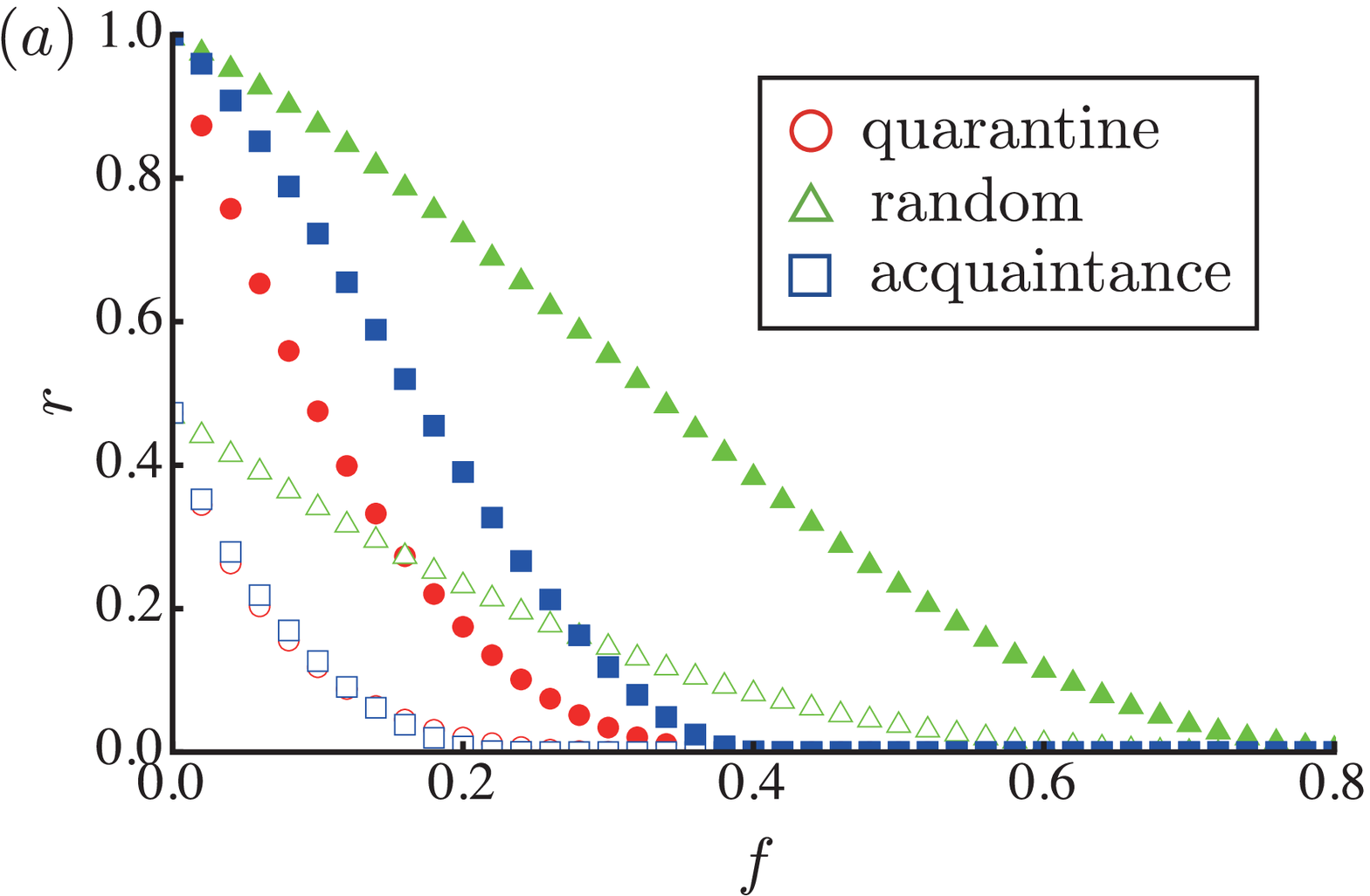}
\includegraphics[width=80mm]{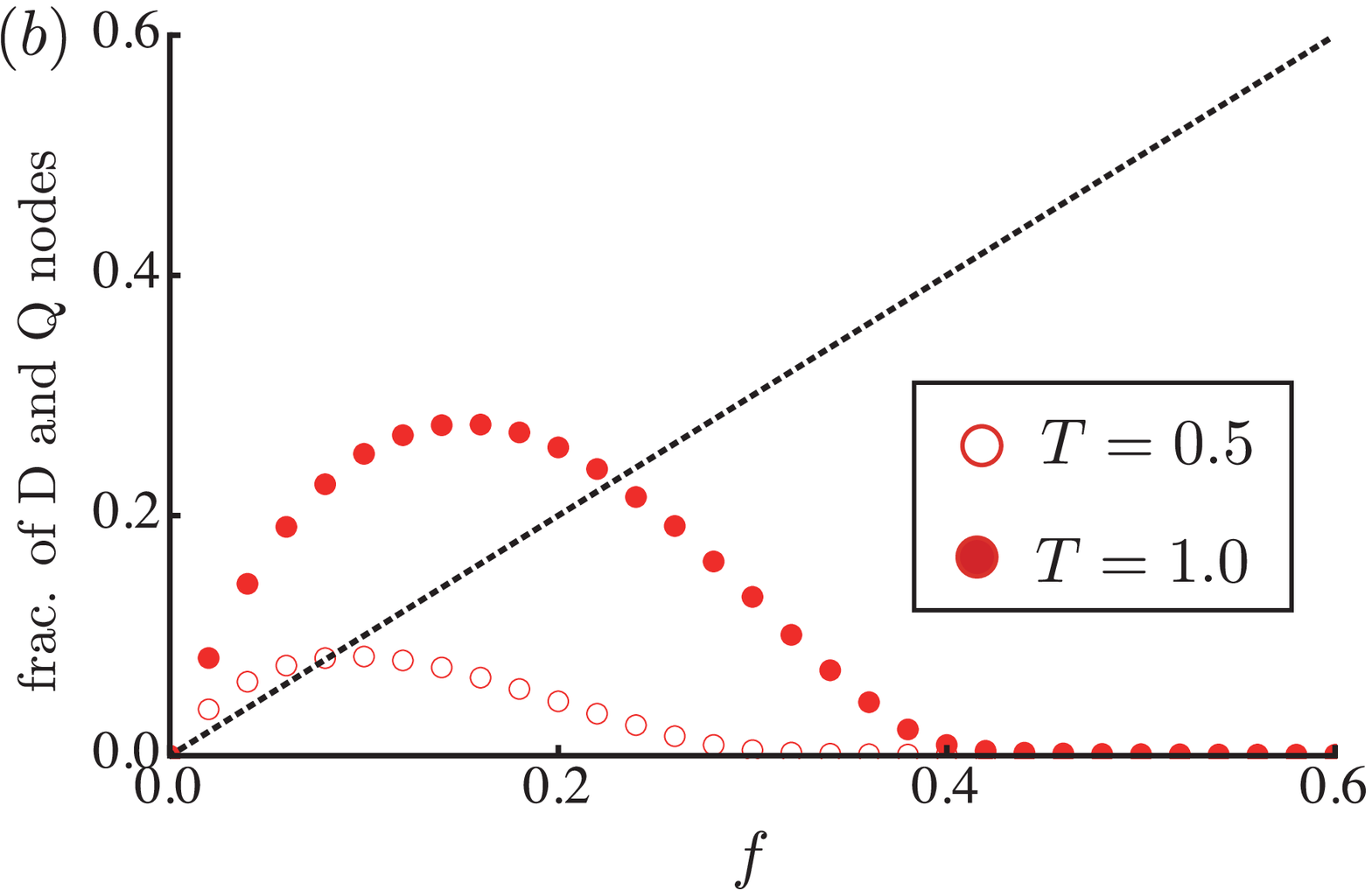}
\end{center}
\caption{
The order parameter $r$, as a function of $f$, for the SFNs. Here $T$ is set to $T=0.5$ (open symbols) and $T=1.0$ (full symbols). The red circles, green triangles, and blue squares represent the results for the quarantine measure, the random vaccination, and the acquaintance vaccination, respectively. 
}
\label{fig:fdependence}
\end{figure}
%%%%%%%%%%%%%%%%%%%%%%%%%%%%%%%%%%%

\section{Results}

\subsection{Order Parameter}

To test the efficiency of the quarantine measure to suppress epidemics, Monte Carlo simulations are performed for the SIR model with the quarantine measure in the two following typical networks: the uncorrelated SFNs with $p_k \propto k^{-2.7} \; (k \ge k_{\rm min}=2)$ that are realized by the configuration model \cite{newman2003structure}, and the RGs with the same mean degree as the SFNs, i.e., $\langle k \rangle \approx 3.844$. The number of nodes is $N=10^5$. The average of quantities at a given $f$ and $T$ is taken over $10^3$ trials $\times$ $10^2$ graph realizations. The detection probability $f$ is set to $f=0.01$ and $0.2$.

Similar simulations are also executed without a control measure, with a random vaccination scheme and with the acquaintance vaccination scheme. In the random vaccination scheme, a fraction of nodes to be vaccinated are randomly selected. In the acquaintance vaccination scheme, a random neighbor of a random node is repeatedly selected for vaccination. In both schemes, nodes are vaccinated prior to the start of an outbreak and the nodes possess perfect immunity such that they never change their state. The fraction of vaccinated nodes is parametrized by $f$ to compare the quarantine and vaccinations. However, it is noted that the actual fraction of D and Q nodes for the quarantine measure does not correspond to $f$.

Figure \ref{fig:orderparameter} plots the mean fraction of the R nodes, $r$, as a function of $T$. With respect to the RGs (Fig.~\ref{fig:orderparameter} (a)), the quarantine measure outperforms the vaccination schemes in terms of reducing the number of R nodes. This is also applicable for the SFNs (Fig.~\ref{fig:orderparameter} (b)). With respect to the SFNs, there are hubs with numerous neighbors through which many chances of becoming infected and infecting other nodes exist. In the quarantine measure, hubs do not appear to leverage their spreading abilities because they can be easily isolated. That is, a hub is quarantined if only one of its numerous neighbors is detected. 

The fraction of nodes infected once may be adopted as the order parameter for the quarantine measure instead of the fraction of the R nodes. The difference is that the order parameter $r$ does not include the D nodes and Q nodes who were already infected when they were isolated. Nevertheless, the superiority of the quarantine measure is almost unchanged even when such an order parameter is adopted (not shown).

Figure~\ref{fig:fdependence}(a) plots the $f$ dependence of $r$ for the SFNs when $T$ is large ($T=0.5$ and 1.0). The quarantine measure succeeds in reducing outbreak size when compared with that of other vaccination schemes. Figure~\ref{fig:fdependence}(b) plots the mean number of isolated nodes (nodes in the D or Q state) as a function of $f$. When $f$ is not too small, the number of isolated nodes is fewer than that of vaccinated nodes because epidemics can be immediately detected at an early stage and eradicated by isolations. Specifically, the quarantine measure can contain epidemics even with $T=1.0$ if $f>f_c \simeq 0.4314$, where $f_c$ is given by Eq.~(\ref{eq:Tc}) as derived below.

%%%%%%%%%%%%%%%%%%%%%%%%%%%%%%%%%%%
\begin{figure}
%[!b]
\begin{center}
\includegraphics[width=80mm]{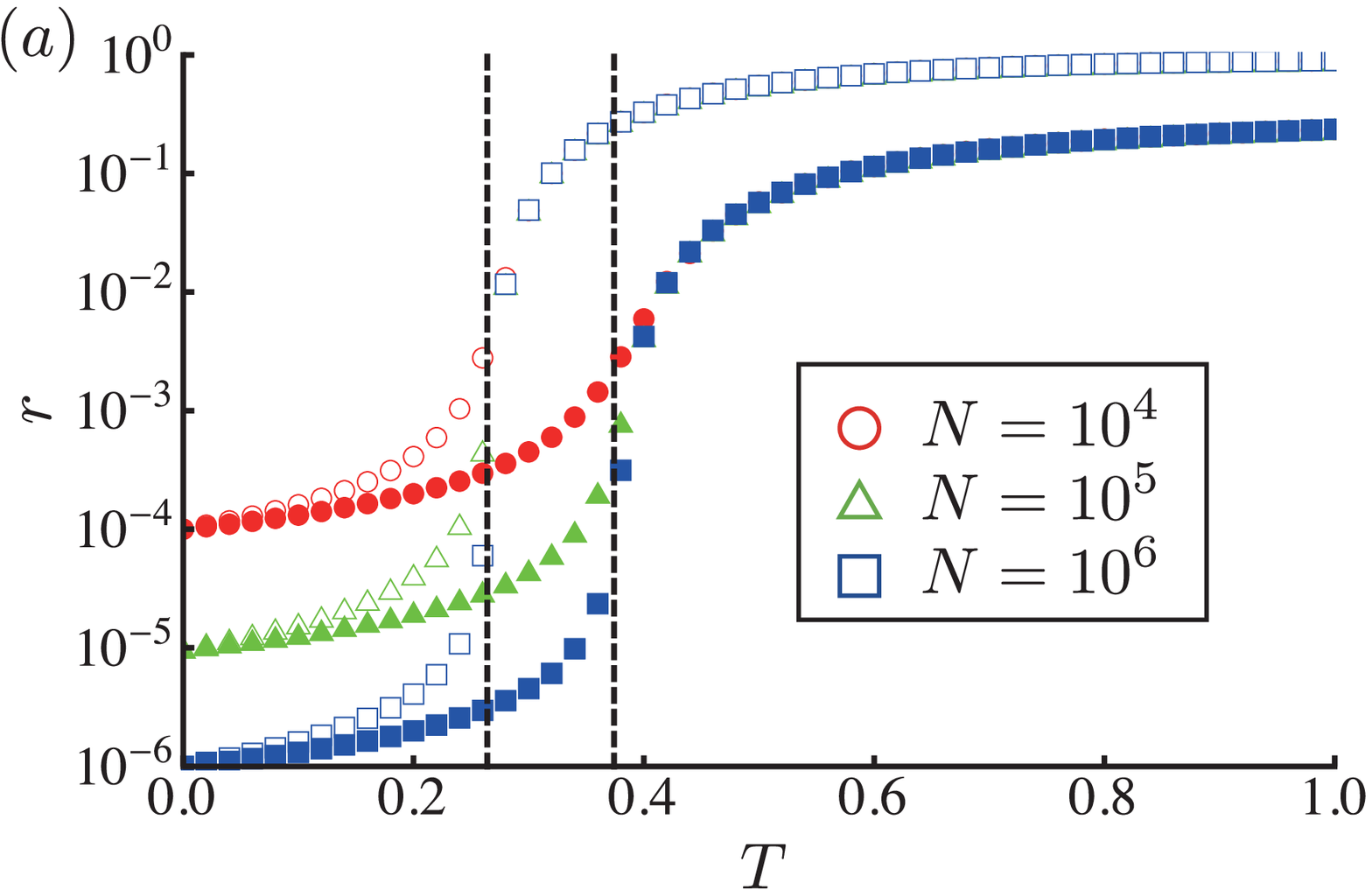}
\includegraphics[width=80mm]{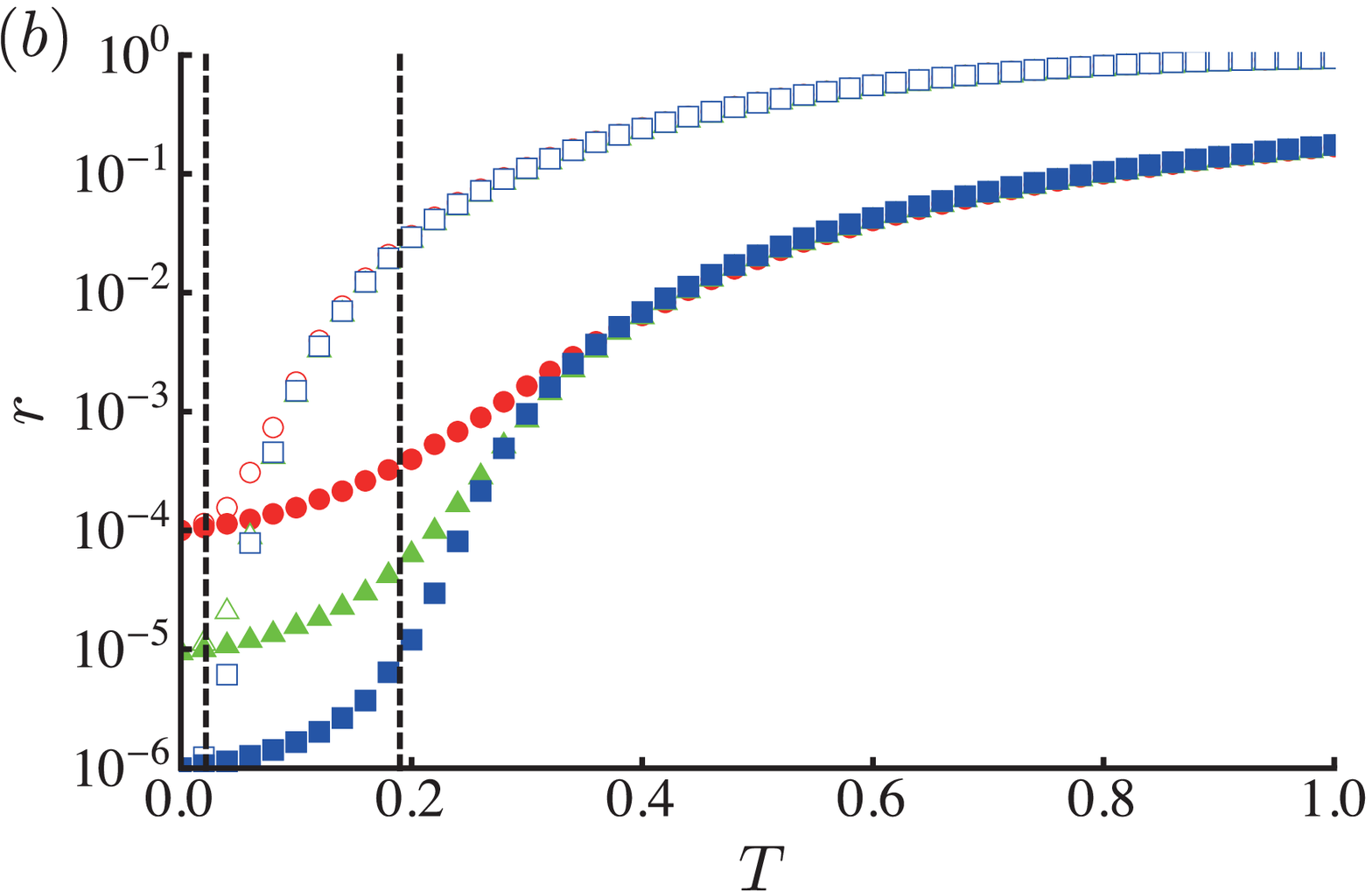}
\end{center}
\caption{
Logarithmic plot of the order parameter, $r$, for (a) the RGs and (b) the SFNs with $N=10^4$ (red circles), $10^5$ (green triangles), and $10^6$ (blue squares). The open symbols and full symbols represent the results of $f=0.01$ and $0.2$, respectively. Two vertical lines represent the epidemic threshold $T_c$ of $f=0.01$ (left) and $f=0.2$ (right) as given by Eq.~(\ref{eq:Tc}).
}
\label{fig:sizedependence}
\end{figure}

\begin{figure}
%[!b]
\begin{center}
\includegraphics[width=80mm]{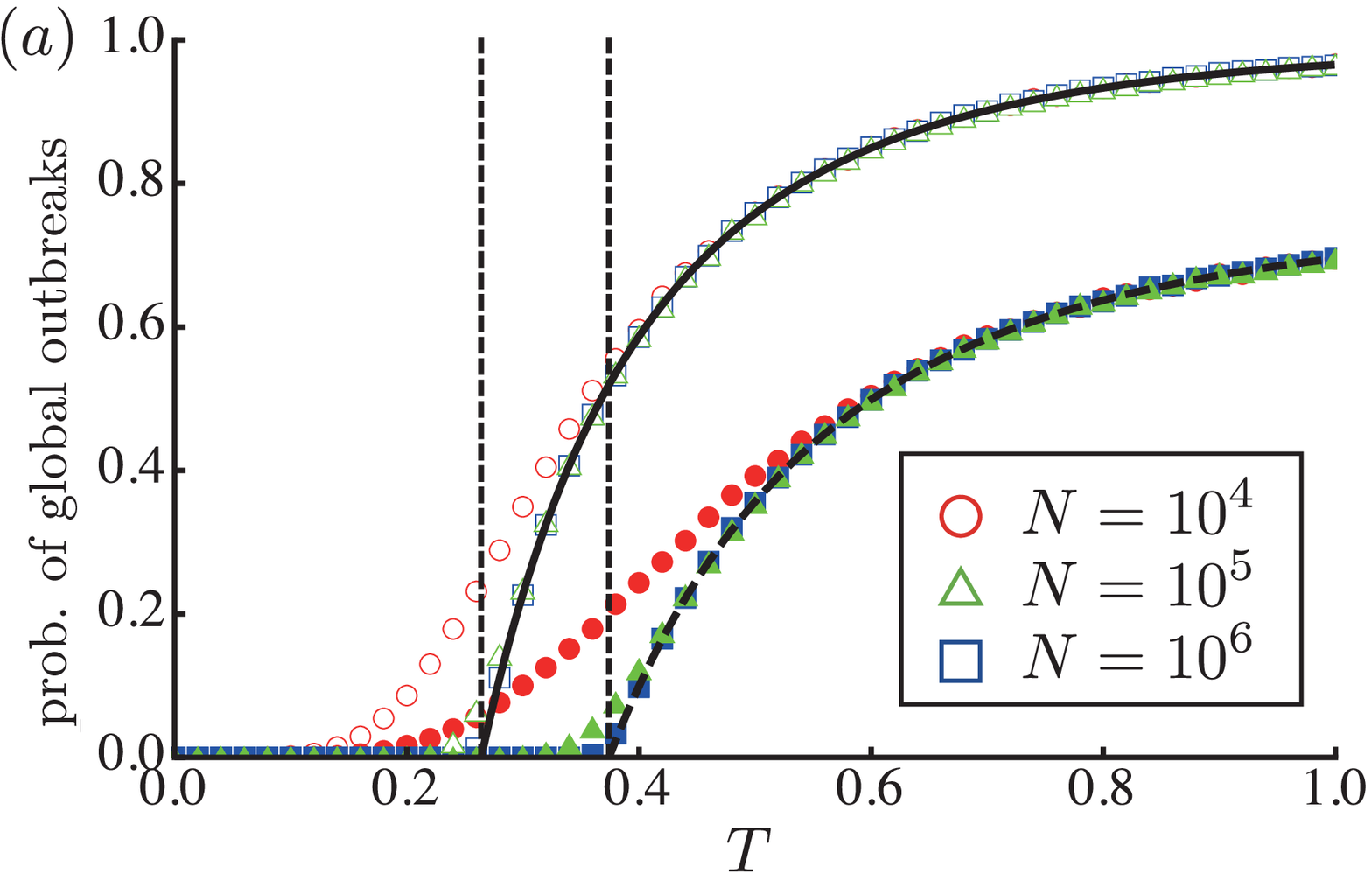}
\includegraphics[width=80mm]{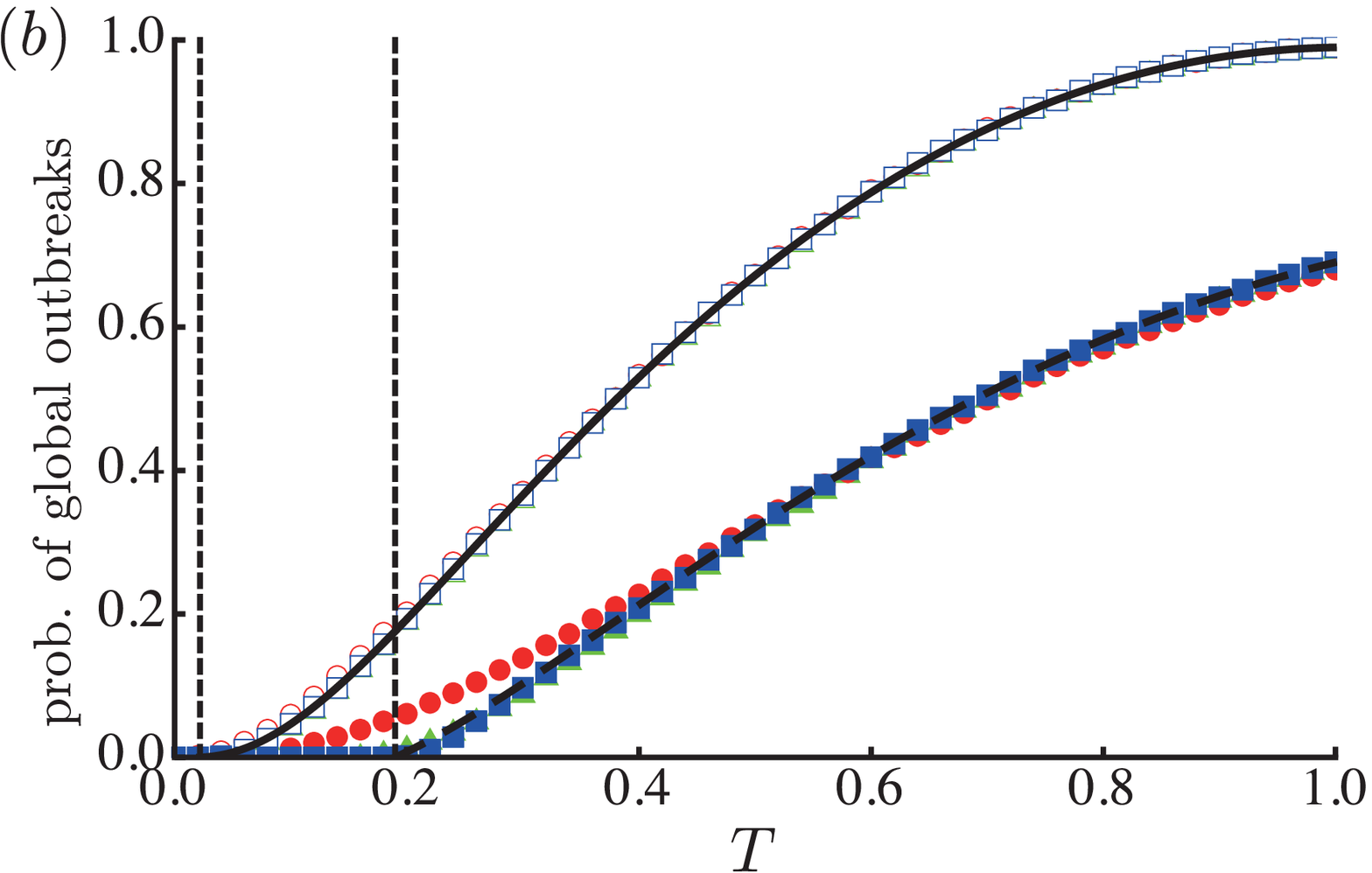}
\end{center}
\caption{
The probability that a single infected node induces a global outbreak for (a) the RGs and (b) the SFNs with $N=10^4$ (red circles), $10^5$ (green triangles), and $10^6$ (blue squares). The detection probability is set to $f=0.01$ (open symbols) and $0.2$ (full symbols), in which symbols denote the fraction of samples such that the fraction of R nodes exceeds 1 \%. The solid and dashed lines represent the probability of global outbreaks (\ref{eq:Prob}) for $f=0.01$ and $0.2$, respectively. Two vertical lines represent the epidemic threshold $T_c$ of $f=0.01$ (left) and $f=0.2$ (right), as given by Eq.~(\ref{eq:Tc}).
}
 \label{fig:probability}
 \end{figure}
%%%%%%%%%%%%%%%%%%%%%%%%%%%%%%%%%%%

\subsection{Epidemic Threshold, Occurrence Probability of Global Outbreaks, and Phase Diagram}

The epidemic threshold and the occurrence probability of global outbreaks are derived by using a generating function formalism \cite{newman2002spread}. An infinitely large uncorrelated network with degree distribution $p_k$ is assumed. The generating function $G_0(x)$ for the degree distribution $p_k$ is defined as follows: 
\begin{equation}
G_0(x) = \sum_{k=k_{\rm min}}^\infty p_k x^k.
\end{equation}
A node reached by following a randomly selected edge is considered. This node has other $k-1$ neighbors, whose number is termed as the excess degree, with probability $q_{k-1} = k p_k/\langle k \rangle$. The generating function $G_1(x)$ for the excess degree distribution $q_k$ is given as follows: 
\begin{equation}
G_1(x) = \sum_{k=k_{\rm min}}^\infty q_{k-1} x^{k-1}=\sum_{k=k_{\rm min}}^\infty \frac{k p_k}{\langle k \rangle} x^{k-1}.
\end{equation}

This is followed by considering an early stage of an outbreak under the quarantine measure. When an I node is adjacent to an S neighbor, then the state of the neighbor remains as S with probability $1-T$, becomes I with probability $(1-f)T$, and becomes D with probability $fT$. An I node with $k$ neighbors is changed to Q and subsequent transmissions are not performed, when one of the neighbors becomes D. During transmissions between an I node and $k$ S neighbors, the probability that the $k'$th neighbor ($1 \le k' \le k$) becomes D is $fT(1-T+(1-f)T)^{k'-1}$ and the probability that no neighbors become D is $(1-T+(1-f)T)^{k}$. Therefore, the generating function $F_0(x)$ for the probability distribution of the number of newly infected neighbors from a randomly chosen I node is as follows: 
\begin{eqnarray}
F_0(x)
&=&\sum_{k=k_{\rm min}}^\infty p_k \Big[ \sum_{k'=1}^{k} fT (1-T+(1-f) T x)^{k'-1} + (1-T +(1-f) T x)^{k} \Big] 
\nonumber \\
&=& fT \frac{1-G_0(1-T + (1-f) T x)}{1-(1-T + (1-f) T x)}+G_0(1-T + (1-f) T x). 
\end{eqnarray}
Similarly, $F_1(x)$ denotes the generating function for the probability distribution of the number of newly infected neighbors from an I node that is reached by following a randomly chosen edge as follows: 
\begin{eqnarray}
F_1(x)
&=&\sum_{k=k_{\rm min}}^\infty \frac{k p_k}{\langle k \rangle} \Big[ \sum_{k'=1}^{k-1} fT (1-T+(1-f) T x)^{k'-1} + (1-T +(1-f) T x)^{k-1} \Big] 
\nonumber \\
&=& fT \frac{1-G_1(1-T + (1-f) T x)}{1-(1-T + (1-f) T x)}+G_1(1-T + (1-f) T x).
\end{eqnarray}
The infections spread only if the mean offspring number $F_1'(1)$ exceeds one, and thus the epidemic threshold $T_c(f)$ is given by the following condition: 
\begin{equation}
F_1'(1)=1 \iff \frac{1-f}{f}(1-G_1(1-fT_c))=1. \label{eq:Tc}
\end{equation}
In the limit $f \to 0$, Eq.~(\ref{eq:Tc}) reduces to a known result for the original SIR model as
\begin{equation}
T_c(0)=\frac{1}{G_1'(1)}=\frac{\langle k \rangle}{\langle k(k-1) \rangle}. \label{eq:TcSIR}
\end{equation}

The probability that a seed induces a global outbreak for $T>T_c(f)$ is also derived. Let $P_s$ denote the probability that a seed induces an outbreak in which $s$ nodes were once infected, and $Q_s$ the probability that a node infected by another node causes the infections of $s$ nodes. Then, the generating functions for $P_s$ and $Q_s$ are given as $H_0(x)=\sum_s P_s x^s$ and $H_1(x)=\sum_s Q_s x^s$, respectively. The recursive relations for $H_0(x)$ and $H_1(x)$ are as follows: 
\begin{equation}
H_0(x)=x F_0(H_1(x)) \quad {\rm and} \quad H_1(x) = x F_1(H_1(x)).
\end{equation}
Furthermore, $H_0(1)=\sum_s P_s$ denotes the probability that an epidemic that begins from a seed terminates with finite infections, and the occurrence probability of global outbreak is expressed as 
\begin{equation}
1-H_0(1)=1-F_0(v), \label{eq:Prob}
\end{equation}
where $v$ is the solution of 
\begin{equation}
v=F_1(v). \label{eq:Prob2}
\end{equation}

To check the aforementioned estimate, the $N$ dependence of the order parameter $r$ is considered for the RG (Fig.~\ref{fig:sizedependence} (a)) and the SFN (Fig.~\ref{fig:sizedependence} (b)). Monte Carlo simulations confirm that $r$ of $N$ nodes approaches zero for $T<T_c$ with increasing $N$. Figure~\ref{fig:probability} plots the probability of global outbreaks given by Eq.~(\ref{eq:Prob}). In the Monte Carlo simulations, a fraction of samples such that the fraction of R nodes exceeds 1 \% is regarded as the occurrence probability of global outbreaks. For both RGs and SFNs, the analytical results coincide well with the numerical results.

%%%%%%%%%%%%%%%%%%%%%%%%%%%%%%%%%%%
\begin{figure}
%[!b]
\begin{center}
\includegraphics[width=80mm]{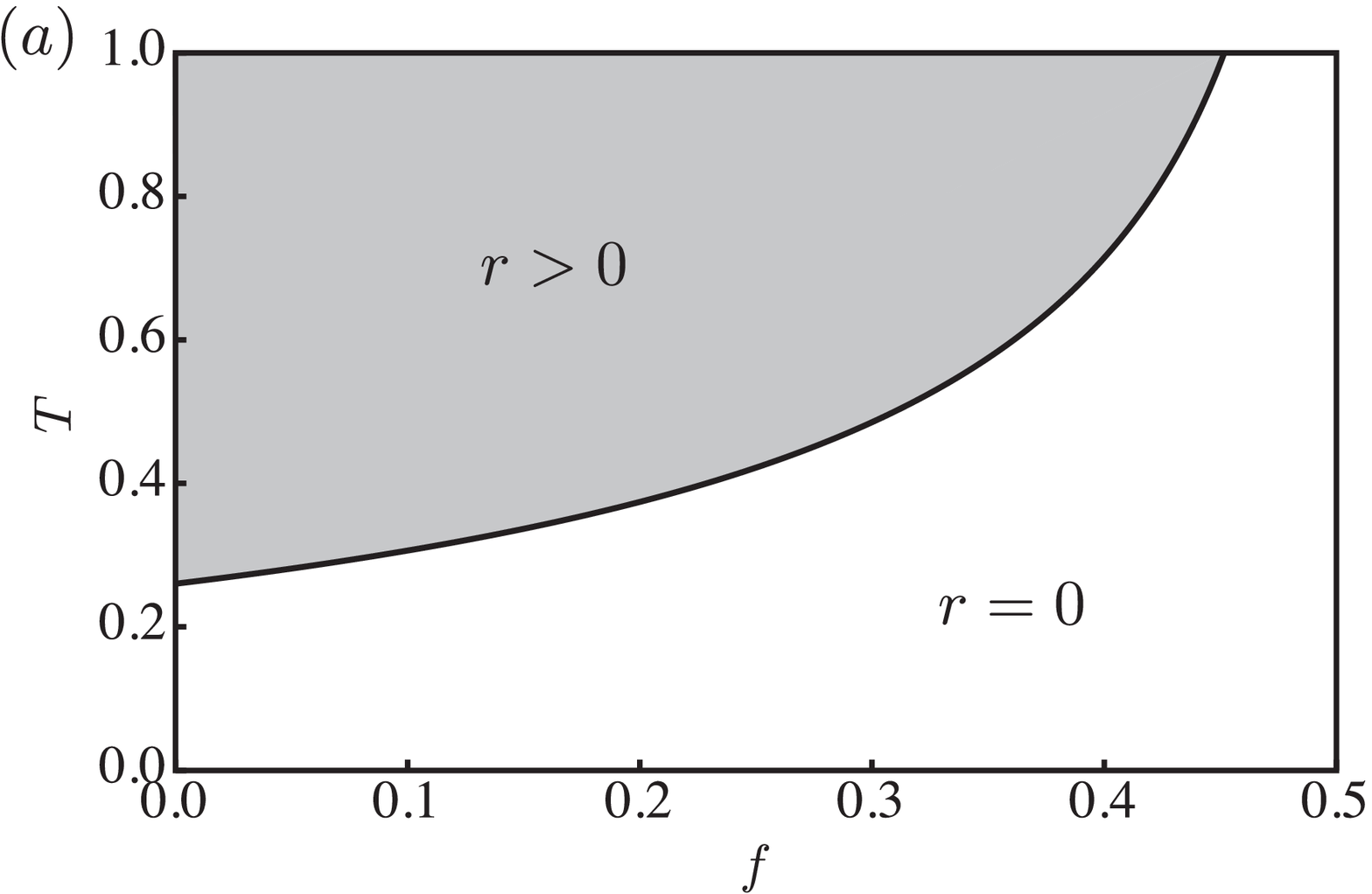}
\includegraphics[width=80mm]{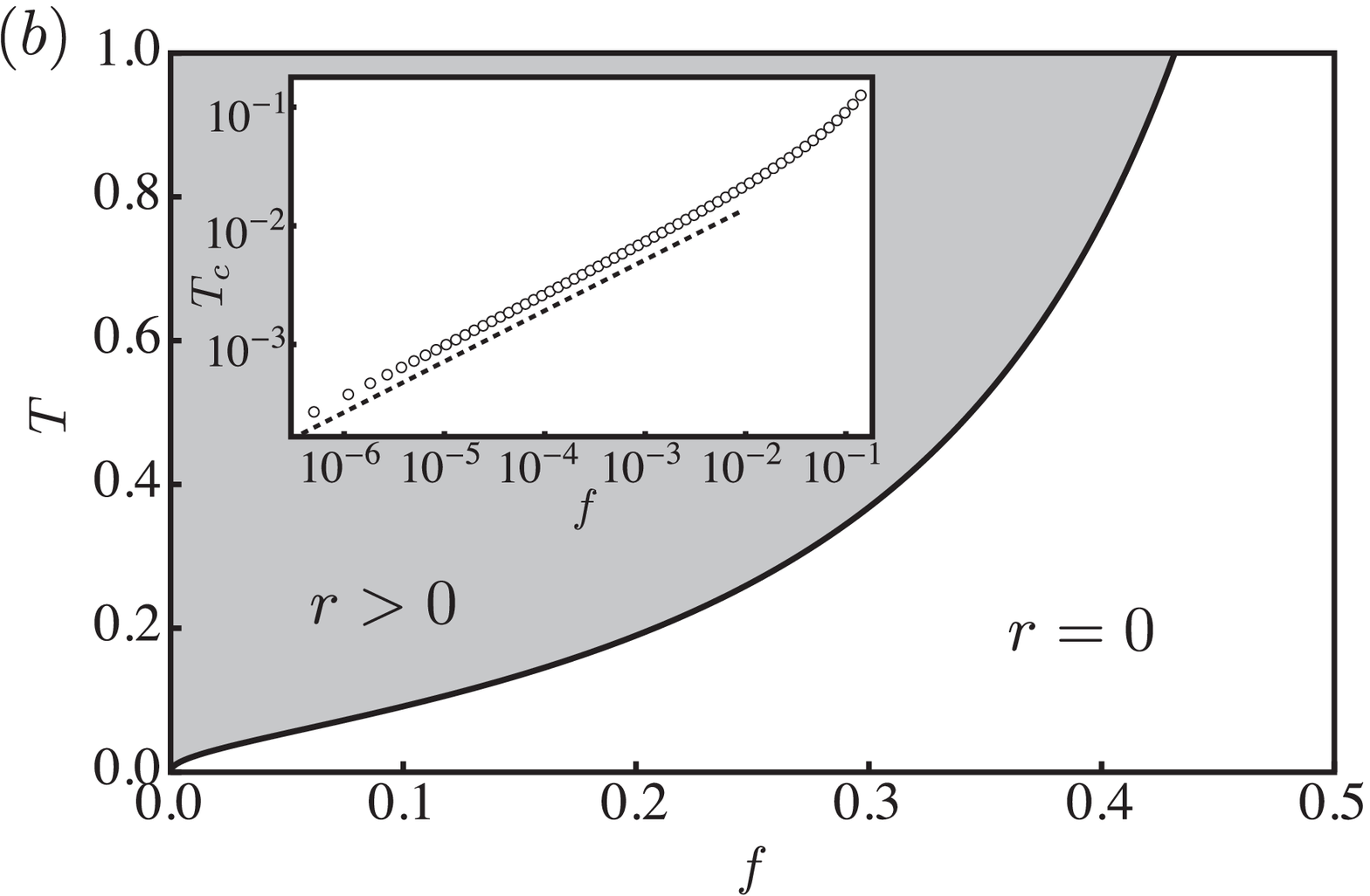}
\end{center}
\caption{
Phase boundary in the $(f, T)$ plane of the SIR model with the quarantine measure on (a) the RGs and (b) the SFNs. The solid lines represent epidemic threshold $T_c(f)$, which is evaluated from Eq.~(\ref{eq:Tc}). Inset of (b): $T_c(f)$ for $f \ll 1$. The dotted line denotes $T_c(f) \propto f^{3/7}$.
}
\label{fig:phasediagram}
\end{figure}
%%%%%%%%%%%%%%%%%%%%%%%%%%%%%%%%%%%

An evaluation of Eq.~(\ref{eq:Tc}) plots the phase boundary in the $(f,T)$ plane as shown in Fig.~\ref{fig:phasediagram} (a) for the RG and Fig.~\ref{fig:phasediagram} (b) for the SFN, respectively. The present quarantine measure is highly effective in increasing the epidemic threshold. Specifically, the quarantine measure increases $T_c(f)$ in the fat-tailed SFNs from zero to a positive value even when the detection probability is small. Expanding Eq.~(\ref{eq:Tc}) for $f \ll 1$ (as shown in Appendix~\ref{sec.App}) results in $T_c(f)$ of uncorrelated SFNs with $p_k \propto k^{-\gamma}$ ($k \ge k_{\rm min}$) as 
\begin{equation}
T_c(f)\simeq\begin{cases}
\alpha_\gamma f^{\frac{3-\gamma}{\gamma-2}}, & 2<\gamma <3, \\
\alpha_3|\log f|^{-1}, & \gamma=3, \\
T_c(0)+\beta'_\gamma f^{\gamma-3}, & 3<\gamma<4,\\
T_c(0)+\beta_4 f|\log f|, & \gamma=4, \\
T_c(0)+\beta_\gamma f, & 4<\gamma,
\end{cases}
\label{eq:f-dependence}
\end{equation} 
where $a_\gamma$ and $\beta_\gamma (\beta'_\gamma$) denote constants that depend on $\gamma$, and $T_c(0)$ is given by Eq.~(\ref{eq:TcSIR}). Equation (\ref{eq:f-dependence}) shows that $T_c(f)>0$ if $f>0$ even in the fat-tailed SFNs with $\gamma \le 3$, and the deviation $T_c(f)-T_c(0)$ for small $f \ll 1$ obeys a power law of $f$, 
in which the exponent depends on degree exponent of the underlying network, $\gamma$. 

\subsection{Case of Real Networks}

\begin{figure}
%[!b]
\begin{center}
\includegraphics[width=80mm]{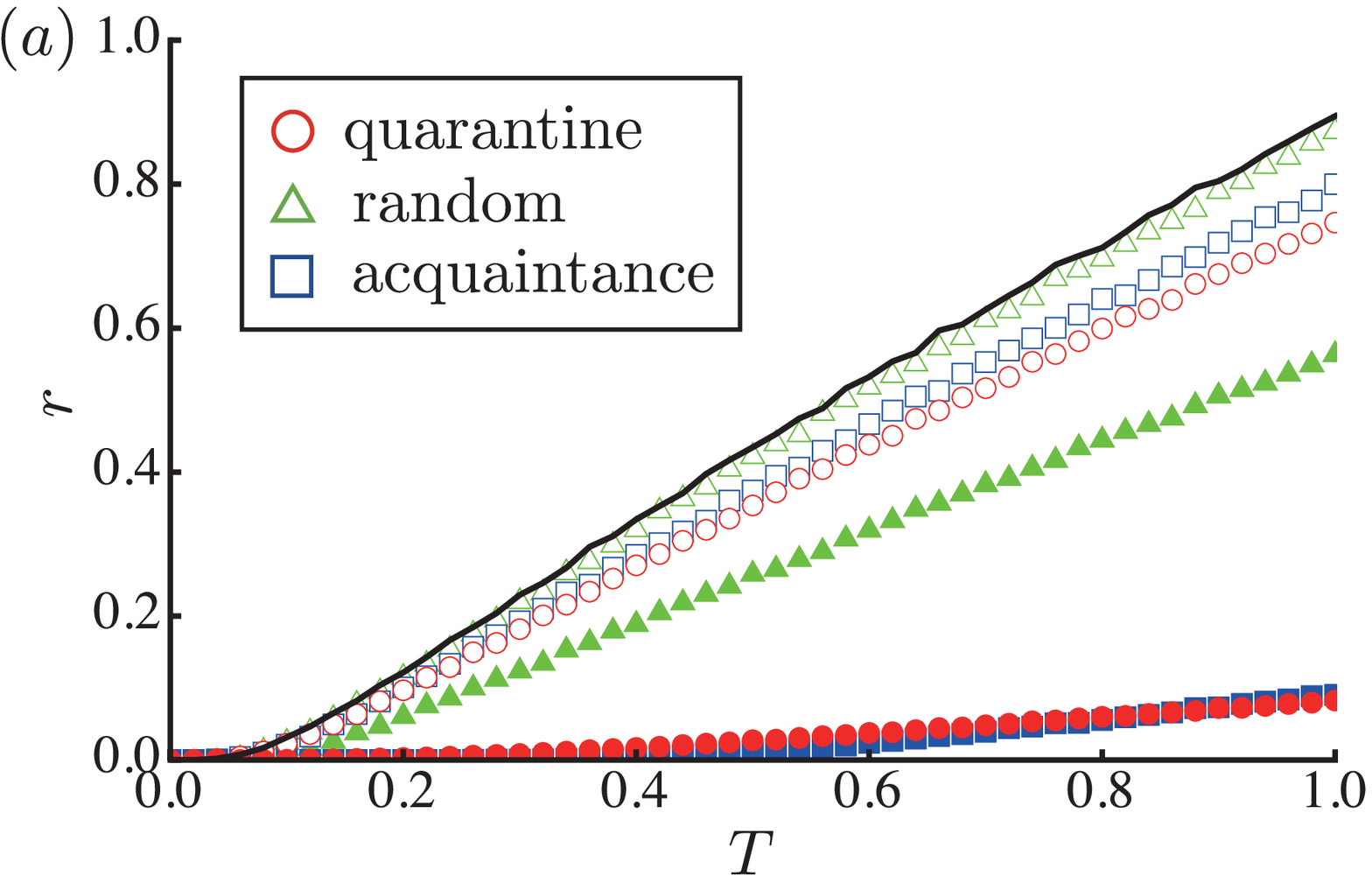}
\includegraphics[width=80mm]{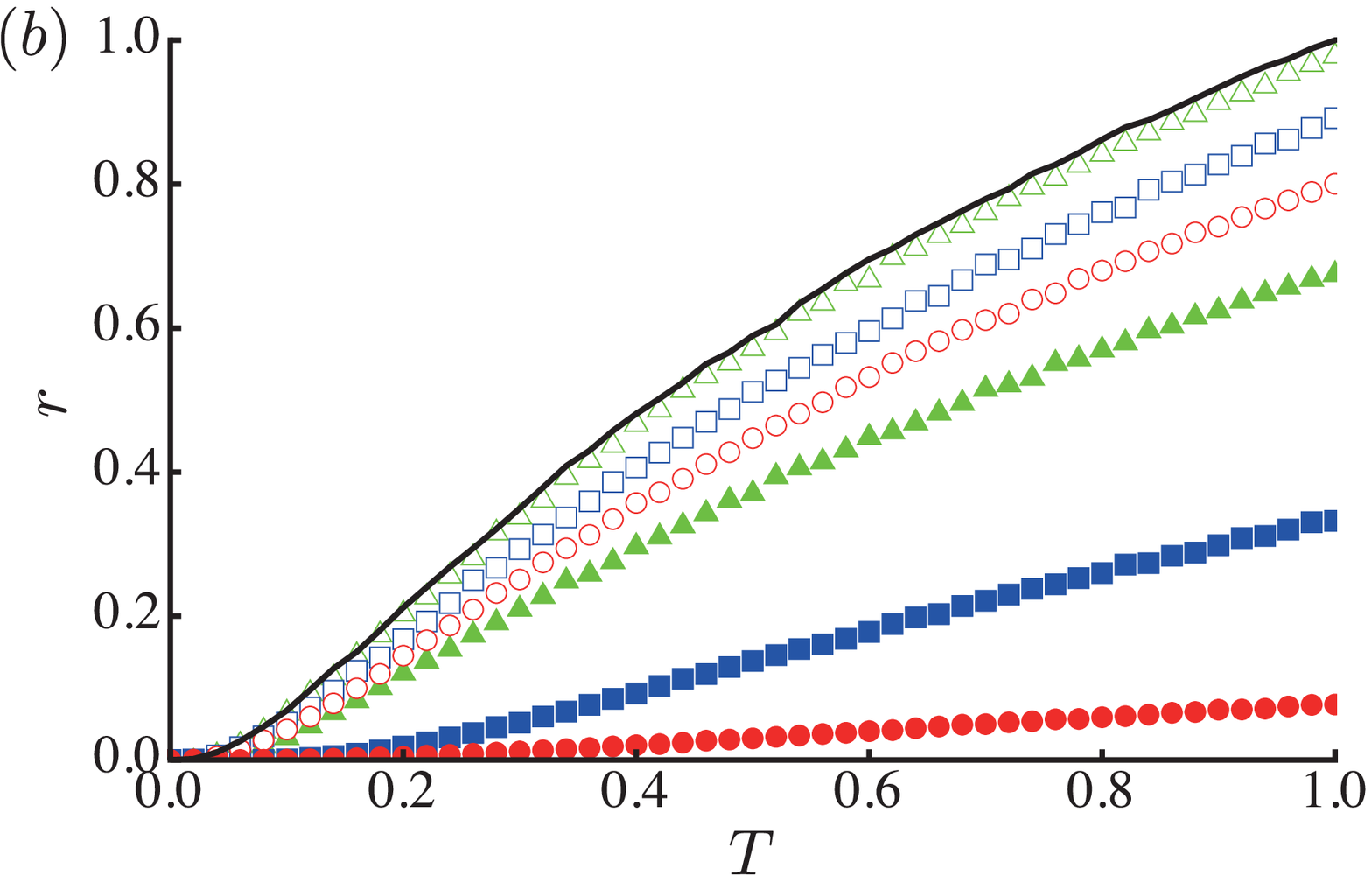}
\end{center}
\caption{
The order parameter $r$, as a function of $T$, for real networks: (a) the sexual contact network between Brazilian prostitutes and sex buyers and (b) the friendship network of Gowalla users. Here $f$ is set to $f=0.01$ (open symbols) and 0.2 (full symbols). The red circles, green triangles, and blue squares represent the results for the quarantine measure, the random vaccination, and the acquaintance vaccination, respectively. The black line represents the result of the original SIR model. The number of nodes is $N=16748$ for the sexual contact network and $N=196591$ for Gowalla network. The data at given $T$ and $f$ are averaged over $10^5$ trials.
}
\label{fig:real}
\end{figure}

The above numerical and analytical results on the efficiency of the present quarantine measure were obtained considering uncorrelated networks. It should be noted that uncorrelated networks do not possess certain important properties of realistic contact networks such as a high clustering coefficient, assortativity, and community structure. However, prompt isolations can effectively contain epidemics in more realistic networks. Figures \ref{fig:real} (a) and \ref{fig:real} (b) show the Monte Carlo results for the two real networks: the sexual contact network between Brazilian prostitutes and sex buyers \cite{rocha2011simulated}, and the friendship network of Gowalla users (Gowalla is a location-based social networking website where users share their locations by checking in) \cite{cho2011friendship}. Since the data of sexual contact network collected by Rocha et al. \cite{rocha2011simulated} constitute a time-ordered list, we consider a time-integrated network, where multiple edges between a node pair are accepted. The mean outbreak size is effectively reduced by the quarantine measure when compared with those of the random and acquaintance vaccinations. Thus, the present quarantine measure is expected to hold effective in real contact networks.

\section{Discussion}

%%%%%%%%%%%%%%%%%%%%%%%%%%%%%%%%%%%
%%%%%%%%%%%%%%%%%%%%%%%%%%%%%%%%%%%

This study involved investigating the manner in which a prompt quarantine measure suppresses epidemics in networks. The proposed simple and ideal quarantine measure assumed that an individual is detected with detection probability $f$ immediately after it becomes infected, and the detected one and its neighbors are promptly isolated. The efficiency of the proposed measure in suppressing the SIR model in the RGs and the uncorrelated SFNs was numerically tested. Monte Carlo simulations indicated that the quarantine measure outperformed the random and acquaintance vaccination schemes with respect to the reduction of the number of R nodes. The generating function formalism for uncorrelated networks was used to obtain the occurrence probability of global outbreaks and the epidemic threshold $T_c$. The equation that derives $T_c$ was expanded to show that the epidemic threshold increases to a positive value even in fat-tailed SFNs given a nonzero detection probability. We also show that the proposed quarantine measure is effective in real contact networks.

The present study assumed an idealized situation, where quarantines can be executed without delay. In practice, there are time lags among one's infection, detection, and quarantine, due to a number of factors (e.g., the time lag to detections by authorities and the time lag to isolations of infected individuals and their neighbors). Realistic epidemiological study must take into account such delay in quarantine measures. Peak et al. \cite{peak2017comparing} investigated the relationship between the effectiveness of quarantine and symptom monitoring, taking into account delay, in containing epidemics and disease dynamics parametrized by seven case-study diseases. They showed that the effectiveness of symptom monitoring and quarantine depends critically on the properties of the infectious disease, such as latent period, infectious period, and transmissibility.

Theoretical studies have been devoted to the effectiveness of different delayed isolations. Pereira and Young \cite{pereira2015control} studied the effectiveness of delayed isolations for infected nodes (not including their neighbors) in controlling susceptible-infected-susceptible epidemics to show that the disease is (not) effectively controlled if the delay in isolating infected nodes is shorter (longer) than a certain critical value. Very recently, Strona and Castellano \cite{strona2017rapid} considered the SIR model with a quarantine measure, having a delay in the early stage of epidemics, and found the rapid decay in its efficiency; if the implementation is not prompt enough, then the quarantines become highly inefficient. For our case, the effectiveness of quarantines is expected to be weakened when a delay among infection, detection, and quarantine is incorporated. For example, the model can be extended to have a delay time $t_{\rm delay}$ for the execution of a quarantine after a node becomes ``detected''. In the simplest setting, an infected node $i$ with degree $k_i$ can try to infect further $k_{\rm add}=\min(t_{\rm delay}, k_i-k_{\rm D})$ neighbors after its $k_{\rm D}$-th neighbor becomes detected. Monte Carlo simulations for such cases show that the performance of quarantine strategy actually becomes worse with increasing delay time $t_{\rm delay}$ (not shown). The epidemic threshold also decreases as $t_{\rm delay}$ increases and reaches the threshold for the random vaccination with the same value of $f$ when $t_{\rm delay}$ becomes larger than the largest degree $k_{\rm max}$ \footnote{After a short consideration, one finds the generating functions $F_0(x)$ and $F_1(x)$ for the case of $t_{\rm delay} \ge k_{\rm max}$ should be $G_0(1-(1-f)T+(1-f)Tx)$ and $G_1(1-(1-f)T+(1-f)Tx)$, respectively. Then, the epidemic threshold is given from the equation, $F_1'(1)=(1-f)T_c \langle k(k-1) \rangle/\langle k \rangle$=1, i.e., $T_c=(1-f)^{-1}\langle k \rangle/\langle k(k-1) \rangle$, which is the epidemic threshold under the random vaccination of $f$.}. 

Further investigation of the effect of delayed quarantines is needed, and in order to incorporate delay time properly it should be discussed by using continuous-time infectious disease models. The epidemic model used in the present study corresponds to the discrete-time SIR model. It is naturally expected that the results can be qualitatively applied in the case of a continuous-time SIR model. It will be an interesting future work to investigate the continuous-time SIR model with delayed quarantines, although the results for our prompt quarantine measure highlight the importance of the speed necessary in detecting and quarantining.

\section*{Acknowledgements}

T.H. thanks to Taro Takaguchi for helpful comments. T.H. acknowledges financial support from JSPS (Japan) KAKENHI Grant Numbers JP15K17716, JP16H03939, and JP26310203. T.H. and K.N. acknowledge financial support from JSPS (Japan) KAKENHI Grant Number JP16K05507.

%%%%%%%%%%%%%%%%%%%%%%%%%%%%%%%%%%%%%%%%%%%%%%%%%%%%%%%%%%%%%

%%%%%%%%%%%%%%%%%%%%%%%%%%%%%%%%%%%%%%%%%%%%%%%%%%%%%%%%%%%%%

\appendix

\section{Derivation of Power-law Behaviors for $T_c$ \label{sec.App}}

The epidemic threshold $T_c$ of uncorrelated SFN with $p_k \propto k^{-\gamma}$ ($k \ge k_{\rm min}$) is considered when $f \ll 1$. For the purposes of convenience, it is assumed that $k_{\rm min}=1$. In this case, the generating function for the excess degree distribution $G_1(x)$ is
\begin{equation}
G_1(x)=\sum_{k=1}^\infty \frac{kp_k}{\langle k\rangle}x^{k-1}=\frac{1}{\zeta(\gamma-1)}\sum_{k=1}^\infty\frac{x^{k-1}}{k^{\gamma-1}},
\end{equation}
where $\zeta(s)$ denotes the Riemann $\zeta$ function, $\zeta(x)=\sum_{k=1}^\infty k^{-x}$.

To discuss the $f$ dependence of $T_c$ for $f \ll 1$, a few properties of $G_1(x)$ are first listed. A function with an integral representation 
\begin{equation}
\phi_s(x)=\int_0^\infty \frac{u^{s-1}}{\e^u-x}\d u=\Gamma(s)\sum_{k=1}^\infty \frac{x^{k-1}}{k^s}
\end{equation}
is introduced such that $G_1(x)$ is expressed as
\begin{equation}
G_1(x)=\frac{\phi_{\gamma-1}(x)}{\zeta(\gamma-1)\Gamma(\gamma-1)}.
\end{equation}
The function $\phi_s(x)$ is defined for $|x|<1$ and $s>0$ and is related to the polylogarithm function, 
$\Li_s(x)=\sum_{k=1}^\infty x^k k^{-s}$, as follows: 
\begin{equation}
x\phi_s(x)=\Gamma(s)\Li_s(x).
\end{equation}
The Taylor expansion of $\phi_s(x)$ is considered. The $n$th derivative with respect to $x$, denoted as $\phi_s^{(n)}(x)$, is expressed as follows: 
\begin{equation}
\phi_s^{(n)}(x)=\frac{\d^n \phi_s(x)}{\d x^n}=n!\int_0^\infty \frac{u^{s-1}}{(\e^u-x)^{n+1}}\d u.
\end{equation}
It should be noted that $\displaystyle \phi_s^{(n)}(1)=\lim_{x\rightarrow 1-}\phi_s^{(n)}(x)$ exists as long as $s>n+1$. The relation 
\begin{eqnarray}
\frac{1}{\e^u-x-\delta}=\frac{1}{\e^u-x}+\frac{\delta}{\e^u-x}\frac{1}{\e^u-x-\delta}
=\sum_{m=0}^{n-1}\frac{\delta^m}{(\e^u-x)^{m+1}}+\frac{\delta^{n}}{(\e^u-x)^{n}}\frac{1}{\e^u-x-\delta}
\end{eqnarray}
is used to obtain the Taylor expansion formula for $\phi_s(x)$ as
\begin{align}
\phi_s(x+\delta)=&\sum_{m=0}^{n-1}\frac{\delta^m}{m!}\phi_s^{(m)}(x)+R_s^{(n)}(x,\delta),\\
R_s^{(n)}(x,\delta)=&\int_0^\infty \frac{u^{s-1}\delta^{n}}{(\e^u-x)^{n}(\e^u-x-\delta)}\d u.
\end{align}
By setting $\delta=-\epsilon<0$ and taking the limit $x\rightarrow 1-$, the above expansion gives
\begin{align}
\phi_s(1-\epsilon)=&\sum_{m=0}^{n-1}\frac{(-\epsilon)^m}{m!}\phi_s^{(m)}(1)+(-1)^nr_s^{(n)}(\epsilon),\\
r_s^{(n)}(\epsilon)=&\int_0^\infty \frac{u^{s-1}\epsilon^{n}}{(\e^u-1)^{n}(\e^u-1+\epsilon)}\d u,\label{rsn}
\end{align}
which is valid as long as $s>n$. Now we change the integral variable in the r.h.s. as $u=\epsilon v$:
\begin{equation}
r_s^{(n)}(\epsilon)=\int_0^\infty \frac{v^{s-1}\epsilon^{s+n}}{(\e^{\epsilon v}-1)^{n}(\e^{\epsilon v}-1+\epsilon)}\d v.
\end{equation}
When the integer $n$ satisfies $n<s<n+1$, it is possible to evaluate the $\epsilon$ dependence by taking the small $\epsilon$ limit as follows:
\begin{equation}
\lim_{\epsilon\rightarrow 0+}\frac{r_s^{(n)}(\epsilon)}{\epsilon^{s-1}}=\int_0^\infty \frac{v^{s-n-1}}{v+1}\d v.
\end{equation}
The integral of the r.h.s. exists, and it can be concluded that the following expression is applicable:
\begin{equation}
r_s^{(n)}(\epsilon)\sim \epsilon^{s-1},\quad \epsilon\rightarrow0+.
\end{equation}
In the marginal case $s=n+1$, the expression (\ref{rsn}) yields
\begin{equation}
r_{n+1}^{(n)}(\epsilon)\sim \epsilon^{n}|\log\epsilon|,\quad \epsilon\rightarrow0+.
\end{equation}
The above expression provides the evaluation of $G_1(1-fT_c)$ for small $f \ll1$. This is expressed as
\begin{equation}
G_1(1-\epsilon)\simeq\begin{cases}
1-a'_\gamma \epsilon^{\gamma-2}, & 2<\gamma<3,\\
1-a_3|\log\epsilon|, & \gamma=3,\\
1-a_\gamma\epsilon + b'_\gamma \epsilon^{\gamma-2} & 3<\gamma<4,\\
1-a_4\epsilon + b_4\epsilon|\log\epsilon|, & \gamma=4,\\
1-a_\gamma\epsilon + b_\gamma \epsilon^2 & 4<\gamma,
\end{cases}
\end{equation}
and this leads us to the solution of Eq.~(\ref{eq:Tc}) as Eq.~(\ref{eq:f-dependence}). 

\end{document}